\documentclass[twocolumn,aps,prl,showpacs,amsmath,amssymb]{revtex4}

\usepackage{graphicx}
\usepackage{dcolumn}
\usepackage{bm}

\begin{document}

\title{Quantum Phase Transitions in a Cuprate Superconductor 
Bi$_2$Sr$_{2-x}$La$_x$CuO$_{6+\delta}$}

\author{Yoichi Ando}
 \email{ando@criepi.denken.or.jp}
\author{S. Ono}
\author{X. F. Sun}
\author{J. Takeya}
\affiliation{Central Research Institute of Electric Power Industry, 
Komae, Tokyo 201-8511, Japan}
\author{F. F. Balakirev}
\author{J. B. Betts}
\author{G. S. Boebinger}
\altaffiliation{Present address: National High Magnetic Field Laboratory, 
Tallahassee, Florida 32310.} 
\affiliation{NHMFL, Los Alamos National Laboratory, Los Alamos, New Mexico 87545}
\date{\today}

\begin{abstract}

To elucidate a quantum phase transition (QPT) in
Bi$_2$Sr$_{2-x}$La$_x$CuO$_{6+\delta}$, we measure charge and heat
transport properties at very low temperatures and examine the following
characteristics for a wide range of doping: normal-state resistivity
anisotropy under 58 T, temperature dependence of the in-plane thermal
conductivity $\kappa_{ab}$, and the magnetic-field dependence of
$\kappa_{ab}$. It turns out that all of them show signatures of a QPT at
the 1/8 hole doping. Together with the recent normal-state Hall
measurements under 58 T that signified the existence of a QPT at optimum
doping, the present results indicate that there are two QPTs in the
superconducting doping regime of this material. 

\end{abstract}

\pacs{74.25.Fy, 74.25.Dw, 74.72.Hs}

\maketitle

One of the emerging paradigms in the condensed matter physics is the
ubiquitous {\it competitions} in strongly-correlated systems. For
example, strong correlations in transition-metal oxides such as
manganites and nickelates often result in nanoscale structures
consisting of competing phases \cite{Uehara,NiO}. The competitions
between different ground states sometimes give rise to a quantum phase
transition (QPT) \cite{Sachdev}, which takes place at zero temperature
when quantum fluctuations cause a cooperative ordering of the system to
disappear or change; in fact, the strong correlations in heavy-fermion
systems \cite{Mathur} and in ruthenates \cite{Grigera} are known to be
responsible for a QPT between competing ground states. 
In high-$T_c$ cuprates, competitions between the
kinetic energy, the local exchange interaction, and the long-range
Coulomb interaction produce nanoscale self-organized structure called
stripes \cite{Tranquada,Zaanen,Kivelson}, and it is of significant
current interest that various competing ground states may alternate at
QPTs depending on material parameters and/or external parameters,
causing the electronic properties to be largely governed by the
competitions \cite{Zhang}. 

An important issue associated with the competing ground states is the
quantum criticality, which helps one to sort out the physics in terms of
universal scaling \cite{vdM}. However, the quantum criticality becomes
important only when the competition results in a second-order QPT, while
some microscopic phase separations and associated colossal effects can
happen \cite{Dagotto} when the QPT is first order. Therefore, finding a
QPT in a strongly-correlated system is one thing, and determining
whether there is an associated quantum criticality is quite another. In
the case of cuprates, our understanding of the QPTs and the underlying
orders is still far from satisfactory; in particular, most of the
previous experimental works of cuprates regarding the QPT
\cite{vdM,Aeppli,Valla} just focused on the universal scaling behavior,
but identifying the exact position and the nature of the putative QPT is
probably even more important. To accomplish the latter, one needs to
find a qualitative change in the electronic properties at very low
temperatures as a function of a control parameter (such as doping),
which is normally a formidable task and experiments along this line 
are just emerging \cite{Dagan,Panagopoulos,Balakirev}. 

Very recently, a pulsed magnetic field experiment \cite{Balakirev} found
strong evidence at low temperatures that there is indeed a QPT at
optimum doping in a cuprate superconductor
Bi$_2$Sr$_{2-x}$La$_x$CuO$_{6+\delta}$ (BSLCO). In that work, the doping
dependence of the normal-state Hall coefficient measured under 58-T
magnetic field was found to show a sharp break at optimum doping,
indicative of a phase transition resulting in a dramatic change in the
Fermi-surface states. Notably, the break in the doping dependence became
sharper and sharper with lowering temperature, suggesting that the
observed feature is truly a result of a zero-temperature transition;
incidentally, it was argued that a QPT associated with the
$d$-density-wave order \cite{Sudip1} can produce such a sharp signature
in the Hall coefficient \cite{Sudip2}. 

However, there remains a puzzle in the BSLCO case: In the in-plane
resistivity measurements of BSLCO under 60 T \cite{Ono} that preceded
the Hall measurements, it was found that the insulator-to-metal
crossover, which may also signify a QPT, occurs near the 1/8 (= 0.125)
doping, and this does not fit well with the QPT at optimum doping (0.16
holes per Cu). Therefore, the evidence for the QPT in BSLCO is rather
controversial and a comprehensive picture for the zero-temperature phase
transition(s) in BSLCO needs to be established as a step towards drawing
a general phase diagram of the cuprates. In this work, to elucidate the
zero-temperature phase diagram of BSLCO, we measure various transport
properties at low temperature and carefully search for experimental
signatures of a sharp change in the electronic properties as a
function of doping. It is worthwhile to note that the transport
properties are inherently suited to study the zero-temperature
properties of a system, because they are governed by very low energy
excitations at low temperature.

\begin{figure} 
\includegraphics[clip,width=8.5cm]{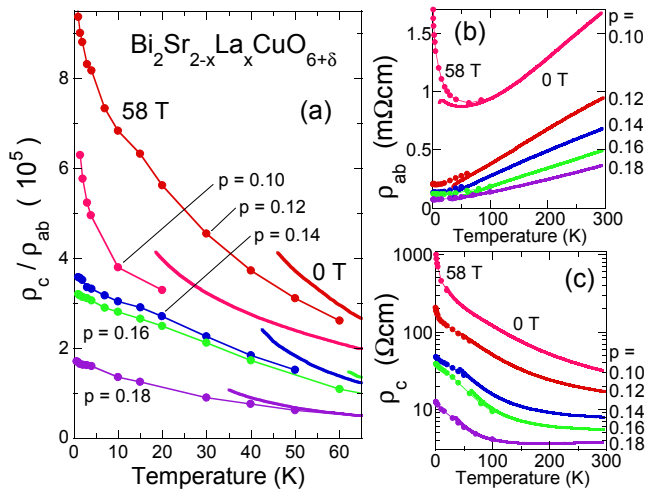}
\caption{Resistivity behavior of BSLCO single crystals. (a) Temperature
dependences of the resistivity anisotropy ratio $\rho_{c}/\rho_{ab}$,
measured in zero field (thick solid lines) and in 58-T field (solid
circles). Note that $\rho_{c}/\rho_{ab}$ remains finite in the
zero-temperature limit for $p \ge 0.14$, while it shows a divergence for
$p \le 0.12$. The data of $\rho_{ab}$ and $\rho_{c}$ used for
calculating $\rho_{c}/\rho_{ab}$ are shown in panels (b) and (c). } 
\end{figure}

As has been noted before \cite{Balakirev,Ono}, BSLCO is an ideal cuprate
system for a systematic study of the low-temperature normal state:
High-quality single crystals can be produced over a wide doping range
\cite{Murayama} and a magnetic field of 60 T is enough to suppress
superconductivity. The hole doping per Cu, $p$, can be controlled
between 0.03 and 0.18 by changing the La content $x$, and the
correspondence between $x$ and $p$ has been sorted out \cite{Hanaki}. In
this work, we essentially employ two experimental techniques we have
been specialized in: resistivity measurements under pulsed magnetic
fields up to 58 T \cite{Ono} and thermal conductivity measurements at
very low temperatures \cite{Takeya,Sun}. Details of each technique are
described in the cited papers. For this work, different sets of samples
\cite{La} are prepared for the measurements of the in-plane resistivity
$\rho_{ab}$ [Fig. 1(b)], out-of-plane resistivity $\rho_c$ [Fig. 1(c)],
in-plane thermal conductivity $\kappa_{ab}$ below 300 mK (Fig. 2), and
the magnetic-field dependence of $\kappa_{ab}$ (Fig. 3); here, the data
for a total of 21 samples are presented, all of which are similar in
size ($\sim$ 1--2 $\times$ 1 $\times$ 0.05 mm$^3$). We concentrate on
identifying the QPT(s) in the superconducting doping regime by looking
at a qualitative change in the electronic properties at very low
temperatures, and it is not the interest of the present study to
determine whether the QPT we find is accompanied by the quantum
criticality. 

\begin{figure}
\includegraphics[clip,width=8.5cm]{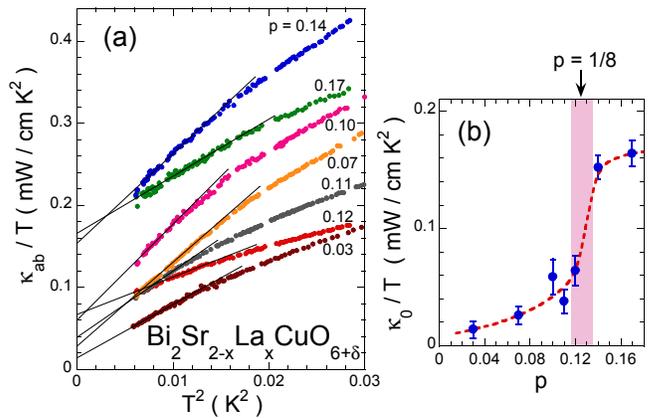}
\caption{Thermal conductivity in 0 T measured in the mK region. 
(a) Plots of $\kappa_{ab}/T$ vs $T^2$ gives the residual
quasiparticle term $\kappa_0/T$ as the zero-intercept of the linear fit
(thin solid lines) to the lowest temperature data. (b) Doping dependence
of $\kappa_0/T$ (sold circles); the dashed curve is a guide to the eyes.
The sudden increase in $\kappa_0/T$ (marked by a shaded band) across $p
\simeq$ 1/8 signals a QPT in the superconducting ground state.}
\end{figure}

The first property we look at is the charge confinement characteristics
\cite{Anderson} in the zero-temperature limit: We measure $\rho_{ab}$ and 
$\rho_c$ in the normal
state by suppressing superconductivity with 58-T pulsed magnetic fields,
and calculate the normal-state anisotropy ratio $\rho_{c}/\rho_{ab}$.
Figure 1(a) shows the temperature dependences of $\rho_{c}/\rho_{ab}$
for $p$ = 0.10 -- 0.18, and Figs. 1(b) and 1(c) show the raw data for
$\rho_{ab}$ and $\rho_c$, respectively. One can easily see in Fig. 1(a)
that there is a qualitative change in the temperature dependence across
$p \simeq$ 0.13; namely, for $p \ge$ 0.14 $\rho_{c}/\rho_{ab}$ hits a
finite value in the zero-temperature limit, while for $p \le$ 0.12
$\rho_{c}/\rho_{ab}$ diverges with lowering temperature. This result
indicates that the characteristics of the charge confinement, which is
one of the most peculiar electronic properties of the cuprates
\cite{Anderson}, changes across $p \simeq$ 1/8 (=0.125). It appears that
for $p <$ 1/8 the charge confinement becomes increasingly more effective
with decreasing temperature, suggesting that the ground state is
strictly two dimensional; on the other hand, since $\rho_{c}/\rho_{ab}$
stays finite for $p >$ 1/8, the ground state can be viewed as an
anisotropic three-dimensional state on this side, though the anisotropy
is extremely large. Such a change in the effective dimensionality
naturally points to a transformation in the fundamental nature of the
ground state in the zero-temperature limit, and thus is indicative of a
QPT at $p \simeq$ 1/8 in the normal state under high magnetic fields.

\begin{figure*}
\includegraphics[clip,width=13.5cm]{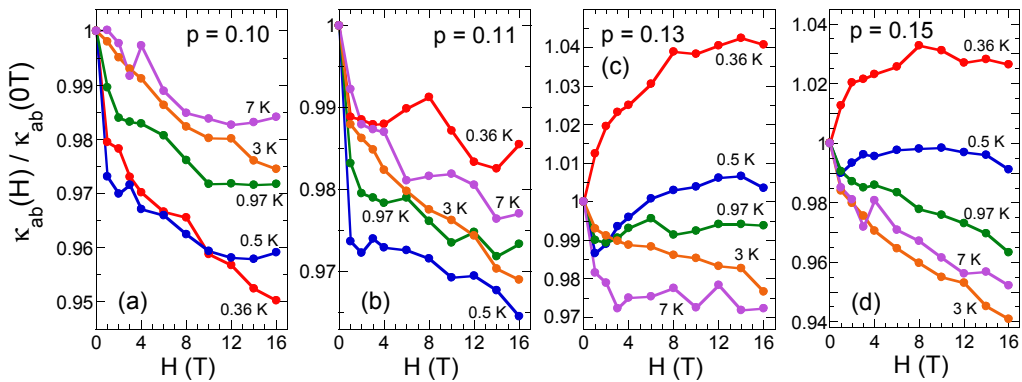}
\caption{Magnetic-field dependences of $\kappa_{ab}$ at various doping levels, 
$p$ = 0.10 (a), 0.11 (b), 0.13 (c), and 0.15 (d). The increase in
$\kappa_{ab}$ with $H$ seen in (c) and (d) at 0.36 K is a standard behavior of
ordinary $d$-wave superconductors where extended quasiparticles are
created by magnetic fields; on the other hand, the decrease in $\kappa_{ab}$ 
with $H$ seen in (a) and (b) even at 0.36 K signifies the
field-induced localization of quasiparticles. 
This qualitative change in the behavior of $\kappa_{ab}(H)$
gives evidence for a QPT at $p \simeq$ 1/8 in the mixed state.} 
\end{figure*}

The second property we look at is the in-plane thermal conductivity
$\kappa_{ab}$ in the mK region, where we can separate the contributions
of phonons and quasiparticles to the heat transport
\cite{Takeya,Taillefer} and therefore the quasiparticle behavior at zero
field (in the superconducting state) can be traced with this tool.
Figure 2(a) shows the plots of $\kappa_{ab}/T$ vs. $T^2$ for 77 -- 170
mK; in these plots, the zero-temperature intercept of the linear fit to
the lowest-temperature data gives the residual quasiparticle term
$\kappa_0/T$, which is a measure of the quasiparticle population at zero
temperature \cite{Takeya,Taillefer}. Remember that in $d$-wave
superconductors the ``impurity band" in the quasiparticle spectrum gives
rise to a finite $\kappa_0/T$, which does not depend on the impurity
concentration but depends on the Fermi velocity and the steepness of the
$d$-wave gap at the nodes in the clean limit \cite{Durst}. The slopes of
the linear fits are determined by the phonon contribution, which we have
confirmed to be in the boundary scattering regime (the phonon mean free
paths \cite{Takeya,Taillefer} estimated from the slopes are consistent
with our crystal dimensions within a factor of 0.4 -- 1.3). In Fig.
2(a), although the range of the data over which we can apply linear
fitting is rather limited, we can determine $\kappa_0/T$ with a certain
error bar; for example, the $\kappa_0/T$ values for $p$ = 0.03, 0.07,
and 0.14 are obtained within an error of $\pm$0.01 mW/cmK$^2$
\cite{error}. As shown in Fig. 2(b), the $p$-dependence of $\kappa_0/T$
shows a jump across $p \simeq$ 1/8 and this jump is much larger than our
error bar. It is useful to note that the $\kappa_0/T$ value for $p$ =
0.14 and 0.17 is $\sim$0.16 mW/cmK$^2$, which is essentially the same as
the values obtained for other cuprates at optimum doping
\cite{Sutherland}; therefore, the superconducting state of BSLCO near
optimum doping is considered to be canonical. What is unusual is the
small values of $\kappa_0/T$ for $p < 1/8$, which is not easily
understood \cite{Takeya,Sutherland} within the standard theory for
$d$-wave superconductors \cite{Durst}. (Similar anomaly in the behavior
of $\kappa_0/T$ was also reported for La$_{2-x}$Sr$_x$CuO$_4$ (LSCO)
\cite{Takeya,Sutherland}.) Although the exact reason for the small
$\kappa_0/T$ value is not known, it is probably related to the
``insulating" nature of the normal state under high magnetic field
\cite{Ono} and is possibly a result of some novel localization effects
\cite{Fisher,Hirschfeld}. In any case, the jump of $\kappa_0/T$ across
$p \simeq$ 1/8 signifies a change in the nature of the superconducting
state at zero temperature, and thus gives evidence for a QPT in the
superconducting state. 

We further look at the magnetic-field dependence of the low-temperature
thermal conductivity, $\kappa_{ab}(H)$, which has recently been shown
\cite{Sun} to be useful for probing a QPT: In underdoped LSCO, it was
found that the magnetic field induces a new phase where the
superconductivity coexists with a static incommensurate
antiferromagnetism, which can be viewed as a field-induced spin density
wave \cite{Lake}; this novel phase in underdoped LSCO leads to a
field-induced localization of quasiparticles \cite{Sun}, which causes
$\kappa$ to {\it decrease} with $H$ even at subkelvin temperatures
\cite{Sun,Hawthorn}, while $\kappa$ is known to {\it increase} with $H$
at $T \alt 2$ K in cuprates near optimum doping \cite{Sun}. It is
notable that the crossover between the two behaviors occurs abruptly
across optimum doping at very low temperatures in LSCO \cite{Sun}, which
is indicative of a QPT. Figure 3 shows the behavior of $\kappa_{ab}(H)$
of BSLCO, measured at low temperatures down to 0.36 K, for four doping
levels. One can see that the $\kappa_{ab}(H)$ behavior shows a
qualitative change across $p$ = 1/8 at the lowest temperature and this
change is essentially the same as that observed in LSCO across optimum
doping. The increase in $\kappa_{ab}$ with $H$ indicates that extended
quasiparticles (that contribute to the heat transport) are created with
$H$ as in ordinary $d$-wave superconductors
\cite{Kubert,Takigawa,plateau}, while the decrease in $\kappa_{ab}$ with
$H$ for $p <$ 1/8 at subkelvin temperatures is indicative of the
field-induced localization of quasiparticles and suggests a coexistence
of the spin density wave \cite{Sun,Takigawa}. As in LSCO, the sharp
change in the $\kappa_{ab}(H)$ behavior as a function of doping is
indicative of a QPT at $p \simeq$ 1/8 in the mixed state of BSLCO.

The above results show that the low-temperature transport properties
give evidence for a QPT taking place at $p \simeq$ 1/8 in all three
possible states of a type-II superconductor: superconducting Meissner
state, mixed state under intermediate magnetic fields, and the normal
state under high magnetic fields. Therefore, the present set of data
adds another QPT to the phase diagram suggested by the normal-state Hall
measurements \cite{Balakirev}, which gave evidence for a QPT at optimum
doping. Furthermore, the present results confirm that the
insulator-to-metal crossover observed in the previous $\rho_{ab}$
measurements \cite{Ono} was indeed due to a QPT. Based on these results,
the phase diagram concluded for BSLCO can be summarized as follows: The
QPT at $p \simeq 1/8$ (QPT1) separates two regimes, Regime 1 ($p < 1/8$)
and Regime 2 ($1/8 < p < 0.16$). The resistivity anisotropy suggests
that in Regime 1 under high magnetic fields the charge confinement is
strong, which seems to be consistent with the idea that some texturing
of the electrons, such as charge density wave or spin density wave
\cite{Lake}, is fundamentally responsible in Regime 1; such texturing of
the electrons can naturally account for the magnetic-field-induced
localization of quasiparticles signified by the $H$ dependence of
$\kappa_{ab}$. The heat transport behavior cannot be understood by the
standard transport theories for $d$-wave superconductors
\cite{Durst,Kubert} in Regime 1, while the canonical heat transport
behavior is observed in Regime 2. The other QPT at optimum doping (QPT2)
separates Regime 3 ($p > 0.16$) from Regime 2; throughout Regimes 1 and
2 the effective carrier density in the zero-temperature limit measured
by the Hall coefficient shows a linear increase with $T_c$
\cite{Balakirev}, which is reminiscent of the Uemura relation for the
superfluid density \cite{Uemura}, and there appears to be an abrupt
change in the Fermi-surface states at QPT2 \cite{Balakirev}.
Intriguingly, the heat transport properties in the superconducting state
do not give any hint of QPT2. 

The existence of two QPTs in BSLCO probably tells us that the physics of
the cuprates in the superconducting doping regime is governed by
competitions between at least three different ground states. Whatever
the nature of the ground states, it is clear that a number of phases are
competing in the cuprates and therefore a promising model of high-$T_c$
superconductivity must have multiple competing phases as possible ground
states. It is yet to be seen how or whether the competition is
related to the occurrence of superconductivity, but it is intriguing to
see that the cuprates are no exception of the strongly-correlated
systems where ubiquitous competitions govern the essential physics. 

We thank S. Chakravarty, S. A. Kivelson, A. N. Lavrov, and S. Sachdev
for helpful discussions and suggestions. The work at the NHMFL was
supported by NSF and DOE.


\begin{thebibliography}{}
\bibitem{La}
The following is the La content ($p$ value, $T_c$ in Kelvin) of the 
samples used for each measurement.
$\rho_{ab}$ and $\rho_{c}$: 0.84 (0.10, 1.4), 0.66 (0.12, 23), 0.49 (0.14, 31), 
0.39 (0.16, 38), and 0.23 (0.18, 29). 
$\kappa_{ab}(T)$: 1.02 (0.03, 0), 0.92 (0.07, 0), 0.84 (0.10, 1.4), 
0.73 (0.11, 15), 0.66 (0.12, 23), 0.49 (0.14, 31), and 0.30 (0.17, 33).
$\kappa_{ab}(H)$: 0.78 (0.10, 9), 0.70 (0.11, 15), 0.61 (0.13, 22), 
and 0.43 (0.15, 33).
\bibitem{error} 
The $\kappa_0/T$ value depends to some extent on the range of
temperature used for the fitting, and our error bar is an estimate of
the uncertainty coming from the choice of the fitting range. The
temperature range chosen for the fits shown in Fig. 2(a) is the
narrowest for $p$=0.10 (77--105 mK) and the widest for $p$=0.17
(77--134 mK).
\end{thebibliography}
\end{document}